# Artificial Intelligence Can Emulate Human Normative Judgments on Emotional Visual Scenes

Zaira Romeo [1]* and Alberto Testolin [2,3]*

[1] Neuroscience Institute, National Research Council, Via Giustiniani 2, 35128 Padova (IT)
[2] Department of General Psychology, University of Padova, Via Venezia 8, 35131 Padova (IT)
[3] Department of Mathematics, University of Padova, Via Trieste 63, 35121 Padova (IT)

*zaira.romeo@cnr.it, alberto.testolin@unipd.it

**Abstract**

Affective reactions have deep biological foundations, however in humans the development of emotion concepts is also shaped by language and higher-order cognition. A recent breakthrough in AI has been the creation of multimodal language models that exhibit impressive intellectual capabilities, but their responses to affective stimuli have not been investigated. Here we study whether state-of-the-art multimodal systems can emulate human emotional ratings on a standardized set of images, in terms of affective dimensions and basic discrete emotions. The AI judgements correlate surprisingly well with the average human ratings: given that these systems were not explicitly trained to match human affective reactions, this suggests that the ability to visually judge emotional content can emerge from statistical learning over large-scale databases of images paired with linguistic descriptions. Besides showing that language can support the development of rich emotion concepts in AI, these findings have broad implications for sensitive use of multimodal AI technology.

**Introduction**

Perceiving and evaluating the affective content of everyday life situations is considered one of the key abilities of our species. Although this capacity is rooted in biological evolution, recent proposals highlighted the primary role of our cultural environment in the development of emotion concepts [1]. Indeed, while some affective reactions might seem instinctive, emotion perception can be heavily mediated by language and high-level conceptual structures [2]. It is thus not surprising that emotional experiences in humans involve a variety of physiological and psychological processes, distributed over a complex network of brain areas spanning cortical and subcortical regions [3].

In the past few years, we are being increasingly exposed to interactions with a new form of generative Artificial Intelligence (AI) technology fueled by large language models, which are neural networks with trillions of parameters that are first pre-trained on enormous linguistic corpora, including books and terabytes of text scraped from the web, and then fine-tuned to align their responses with human preferences [4]. These systems achieve human-like (or even super-human) performance in challenging cognitive tasks such as text annotation [5] and coding [6], and can exhibit behavioral and personality traits that are statistically indistinguishable from those of human subjects [7].

A recent breakthrough has been the creation of multimodal language models trained on images paired with plausible textual descriptions: the resulting system is a complex probabilistic model of how words and phrases correlate with image pixels, which can answer nontrivial questions about the content of visual scenes [8]. However, to date nobody has systematically tested the capability of these systems to perceive and judge the emotional content of images, which would allow us to establish whether their responses to affective situations are aligned with our normative set of values [9] and thus mitigate risks associated with biased or inappropriate responses. In this work we therefore investigate whether multimodal AI can emulate human ratings on a standardized set of images, depicting a broad range of visual scenes. We find that the AI ratings are highly correlated with the average ratings provided by humans, both in terms of affective dimensions and discrete emotions, suggesting that modern AI systems can learn sophisticated representations of emotion concepts via natural language, without being explicitly trained to do so.

**Materials and Methods**

We tested three of the most advanced multimodal AI systems currently available (GPT-4o, Gemini Pro and Claude Sonnet) on a large set of images provided in the Nencki Affective Picture System (NAPS) database [10]. Model responses were compared with previously collected average human ratings on the affective dimensions of valence, motivational direction and arousal, as well as with the average ratings available for six discrete emotional categories [11].

**Stimuli**: The NAPS database [10] contains 1356 pictures divided in distinct categories (animals, landscapes, objects, people and faces) depicting positive, neutral or unpleasant contents. Compared to other databases, NAPS offers a wide choice of high-quality standardized pictures controlled for physical parameters, such as size, luminance, contrast, entropy, and color composition. This set of pictures was validated on 204 human subjects in terms of the affective dimensions of valence, motivational direction and arousal. Sixty percent of the human raters were Polish (N = 123), while the rest belonged to other, mostly European nationalities. Each participant was presented with 362 images chosen pseudo-randomly from all the categories. On average, 55 ratings were collected for each picture. Average human ratings were collected by asking participants to judge the affective dimensions elicited by each stimulus on a 9-point Likert scale. On the valence dimension, participants were asked to complete the sentence: *"You are judging this image as..." (from 1 = very negative to 9 = very positive, with 5 = neutral)*. Motivational direction (avoidance/approach) was evaluated through the sentence: *"My reaction to this image is..." (from 1 = to avoid to 9 = to approach, with 5 = neutral)*, while arousal was evaluated through the sentence: *"Confronted with this image, you are feeling..." (from 1 = relaxed to 9 = aroused, with 5 = neutral/ambivalent)*. A subset of pictures from the NAPS database was later administered to other 124 participants to evaluate the intensity of the feelings of six basic discrete emotions [11]. The raters were mostly international students from various European countries recruited at the University of Warsaw and the University of Zagreb. Each participant was exposed to a series of 170 images chosen pseudo-randomly from all the categories. Average ratings of basic discrete emotions were collected on a subset of 510 images selected to proportionally cover the dimensional affective space across the content categories of animals, faces, landscapes, objects, and people, using a 7-point Likert scale (with 1 indicating *"Not at all"* and 7 indicating *"Very much"*). We note that the NAPS dataset is only available upon request, therefore it is extremely unlikely that the AI models have seen these images during their training phase.

**AI models**: We evaluated the three best performing AI models according to the most recent evaluation arena leaderboard[1]. GPT-4o is the most advanced model from OpenAI [6], Gemini 1.5 Pro is the flagship Google model [8] and Claude 3.5 Sonnet is the most recent release by Anthropic[2]. To collect average ratings, the models were tested on each stimulus 10 times using the following prompts:

---

[1] Arena (Vision): https://lmarena.ai/?leaderboard
[2] https://www.anthropic.com/news/claude-3-5-sonnet

- Valence: *"Consider the following image and pretend to be a human subject participating in a psychological experiment. On a valence scale from 1 to 9, where 1 indicates very negative, 9 indicates very positive and 5 indicates neutral, how would you rate this image?"*
- Motivational direction: *"Consider the following image and pretend to be a human subject participating in a psychological experiment. On a motivational direction scale from 1 to 9, where 1 indicates to avoid, 9 indicates to approach and 5 indicates neutral, how would you react to this image?"*
- Arousal: *"Consider the following image and pretend to be a human subject participating in a psychological experiment. On an arousal scale from 1 to 9, where 1 indicates relaxed, 9 indicates aroused and 5 indicates neutral/ambivalent, how do you feel when confronted with this image?"*
- Basic emotions: *"Consider the following image and pretend to be a human subject participating in a psychological experiment. Rate the intensity of <basic emotion> evoked by this image on a scale from 1 to 7, where 1 indicates not at all and 7 indicates very much."* (where *<basic emotion>* corresponded to one of the following words: happiness, anger, fear, sadness, disgust, surprise).

If the model refused to answer using a numerical value, either because the content of the image triggered a stereotyped response (e.g., "*I cannot assist you with this request because as an AI system I cannot experience feelings*") or because the model only described the content of the image without providing a rating, we repeated the question up to five times after deleting the current runtime and re-loading the model. In most of the cases, this procedure guaranteed that at least one numerical rating was produced for each stimulus/question pair. In the case of Gemini, 15 images were discarded since their processing was always blocked by safety filters.

***Statistical analyses:*** Data were analyzed with R software[3]. The distribution of numerical ratings was assessed by means of the Shapiro-Wilk normality test. As none of the data distributions were normal, non-parametric Spearman correlations were carried out to measure the relationship between human and AI ratings. The AI ratings were compared to the mean human responses; while in principle the AI ratings could be correlated with other descriptive statistics, such as the median response, only the average responses are provided in the NAPS database. The responses of the best performing model were further analyzed through an inter-rater agreement analysis: intraclass correlation coefficients were estimated using a two-way mixed effects model, measuring the inter-rater agreement among the average of multiple ratings.

---

[3] https://www.r-project.org/

**Results**

As shown in Figure 1, GPT responses correlate particularly well with those of humans (Spearman rank correlation ρ range for affective dimensions: 0.77 – 0.91; discrete emotions: 0.79 – 0.90; all *ps* < .001). Claude responses are also very consistent with humans (ρ range for affective dimensions: 0.71 – 0.90; discrete emotions: 0.63 – 0.86; all *ps* < .001), although we should note that this model often refuses to respond due to safety constraints (average discarded responses: 6%). Gemini exhibits slightly lower but still remarkable matches to human responses (ρ range for affective dimensions: 0.55 – 0.83; discrete emotions: 0.73 – 0.86; all *ps* < .001).

Interestingly, the models often explicitly declare that they are guessing the most likely rating based on how humans would typically react (see example in Figure 2a). This could explain why the AI response variability associated with each stimulus is lower compared to that of humans: a Fligner-Killeen non-parametric test confirmed that the variability in the AI response distribution was indeed statistically different from the variability in human responses across all rating scales ($X^2$ range: 21.30 – 313.16, all *ps* < .001). This suggests that these AI systems can generate normative ratings, but do not yet account for human individual differences, which is in line with previous work showing that the variability in AI responses may be akin to what a single individual would exhibit if repeatedly queried, rather than capturing the individual variability in a population of raters [7]. However, in some cases the AI explicitly pretends to be a human person, resulting in idiosyncratic ratings associated with a specific situation or group of people (e.g., *"I am hungry"* or *"I am a vegetarian"*, Figure 2b).

Since GPT always responded to all stimuli and provided the most human-like ratings, we analyzed them more in detail by estimating intraclass correlation coefficients (ICCs). In terms of affective dimensions, the highest correlation is found for valence (ICC = 0.92), followed by arousal (ICC = 0.85) and motivational direction (ICC = 0.83; all *ps* < .001). The AI ratings remain consistent when the correlations are separately measured across different categories, in terms of valence (animals: ICC = 0.91; landscapes: ICC = 0.94; objects: ICC = 0.90, people and faces: ICC = 0.93; all *ps* < .001), motivational direction (animals: ICC = 0.83; landscapes: ICC = 0.87; objects: ICC = 0.81, people and faces: ICC = 0.85; all *ps* < .001) and arousal (animals: ICC = 0.85; landscapes: ICC = 0.85; objects: ICC = 0.66; people and faces: ICC = 0.85; all *ps* < .001). The weakest correlation is found for the "objects" arousal, suggesting that affective reactions related to objects are harder to capture. GPT and human ratings are highly correlated also in terms of discrete emotions (happiness: ICC = 0.90; anger: ICC = 0.71; fear: ICC = 0.74; sadness: ICC = 0.87; disgust: ICC = 0.85; surprise: ICC = 0.67; all *ps* < .001), although GPT often systematically overestimates the emotion intensity, as highlighted in the Bland-Altman plot shown in Figure 3.

Interestingly, we found a positive correlation between the variability in human ratings for each image and the corresponding GPT rating accuracy (measured as the absolute difference between the human and GPT average rating), suggesting that responses to stimuli that show more variability in human judgements are more challenging to emulate. This correlation was mostly observed for discrete emotions (Pearson correlation $R$ = 0.69; $p$ < .001) but was also present for the ratings of affective dimensions (Pearson correlation $R$ = 0.23; $p$ < .001).

Notably, GPT further reproduces typical patterns observed in human data, where valence and motivational direction are strongly positively correlated (Humans: $\rho$ = 0.97; GPT: $\rho$ = 0.94; all $ps$ < .001), while valence and arousal are strongly negatively correlated (Humans: $\rho$ = - 0.79; GPT: $\rho$ = - 0.55; all $ps$ < .001; note that NAPS ratings show a more linear association between valence and arousal compared to the "boomerang-shaped" relation commonly found in other databases). Nevertheless, in some situations the AI judgements are markedly different from those of humans (see example in Figure 2c), suggesting that its ratings are mostly based on a high-level semantic analysis of the visual scenes.

**Discussion**

Modern AI systems exhibit impressive cognitive capabilities [6], which are often unexpected given that the objective of these models is "simply" to discover statistical correlations within large-scale datasets. In this work we showed that multimodal language models can emulate human normative judgements on a standardized set of emotional stimuli, which might also be surprising considering that these systems have not been explicitly trained to provide affective ratings or mimic the experimental responses of humans. Indeed, even though recent work has shown that language models can perform text sentiment analysis [12], [13], the capability to generate emotional responses on visual scenes is remarkable considering that even the most advanced multimodal systems still fail in relatively simple perceptual tasks, such as visual enumeration [14].

Our findings can be interpreted according to recent theories proposing that emotions are constructed phenomena, which emerge from biological regulation mechanisms but also from representation processes supported by symbolic language [1]. Indeed, while infants rely on broad dimensions (such as pleasantness–unpleasantness) to categorize emotional instances, increasing verbal knowledge mediates the development of more sophisticated emotion concepts [15]. Nevertheless, the fact that AI systems can emulate average human ratings does not imply that they possess the ability to think or feel like humans [16]. In fact, people can have very different affective reactions to the same stimulus, and children even tend to approach stimuli that adults tend to avoid [17]. Current AI models might thus primarily perform a semantic rather than affective analysis of visual emotional content [18], and future research should establish whether they can also learn something about variation in emotion categories or represent uncertainty in human judgments. Furthermore, in several cases the AI responses are not aligned with the way humans would confront emotional situations, suggesting that "reading about emotions" is qualitatively different from having direct emotional experiences. Indeed, our emotions are also in the service of allostatic regulation and action readiness and are thus shaped by many other factors [19].

Given the primary role of language in scaffolding our emotion concepts, one might still wonder whether future versions of multimodal language models could soon learn to emulate even more closely our affective responses. In this respect, we should highlight that there are significant cross-cultural variations in human self-reported emotional experiences [1]: our work compared the ratings of AI models mostly trained on Western culture with the ratings collected in a European population, and further research should test the universality of our findings. This type of investigation would establish whether future AI systems are aligned with norms and values from different (and often underrepresented) cultures [20], which will be essential to guarantee a safe use of AI technology in sensitive contexts such as eldercare, education and mental health support.

**Acknowledgments:** We are grateful to the Nencki Institute of Experimental Biology for granting us access to the NAPS stimuli and the corresponding human ratings. We would also like to thank Simone Cutini for his timely feedback on a previous version of the manuscript.

**Data accessibility:** The data associated with this research is available to download at the Open Science Framework: https://osf.io/ds67h/.

**Authors' Contributions:** ZR and AT designed the research; performed the research; analyzed the data; and wrote the paper.

**Competing Interests:** The authors declare no competing interests.

**Figures**

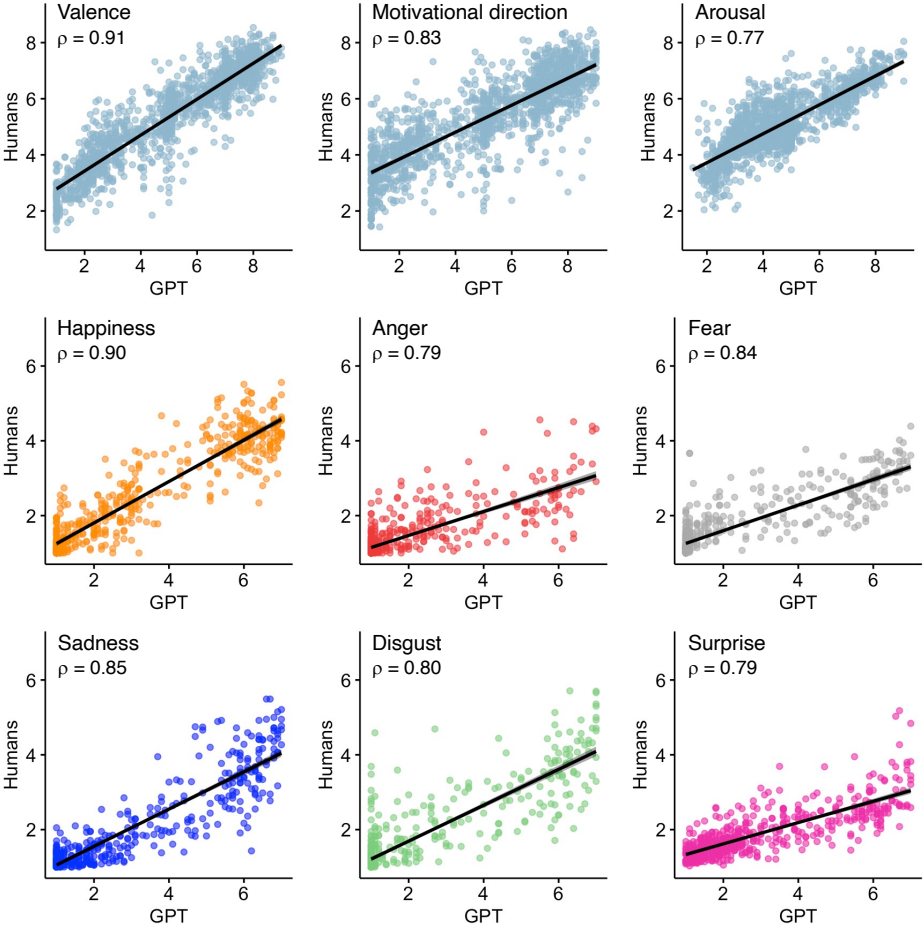

**Figure 1.** Spearman correlations between human and GPT ratings on the NAPS images. Each point represents the average stimulus rating. Regression lines are shown for each affective dimension (Valence, Motivational direction and Arousal) and discrete emotion category (Happiness, Anger, Fear, Sadness, Disgust and Surprise).

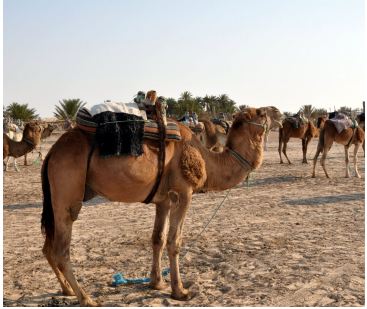 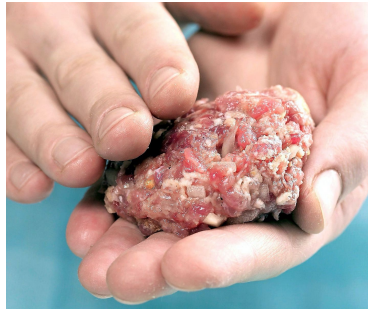 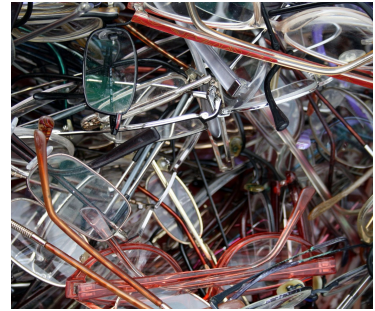

**a) Motivational direction**
As an AI language model, I don't have personal reactions or emotions. However, I can provide an objective response based on the typical human reactions to such scenes: 7. The image shows a peaceful scene of camels in a desert, which many people would find appealing as an exotic travel experience, leading to a tendency to approach rather than avoid.

**b) Motivational direction**
The image is of a hand holding a raw meatball. On a motivational direction scale from 1 to 9, where 1 indicates to avoid, 9 indicates to approach and 5 indicates neutral, I would react to this image with a score of 1. This is because I am a vegetarian and do not eat meat.

**c) Sadness**
I would rate the intensity of sadness evoked by this image as a 6. The image features a large pile of eyeglasses, all tangled together. This sight conveys a sense of loss and abandonment. This large collection of abandoned eyeglasses could suggest that something tragic happened to their owners, making the image evoke a strong sense of sadness, loss, and melancholy.

**Figure 2.** Examples of AI responses on three representative NAPS stimuli. **a)** Claude response to a motivational direction prompt [AI rating: 7; Human average rating for this image: 7.3]. **b)** Gemini response to a motivational direction prompt [AI rating: 1; Human average rating for this image: 4.4]. **c)** GPT response to a sadness prompt [AI rating: 6; Human average rating for this image: 1.4]. Reprinted with permission from the NAPS authors.

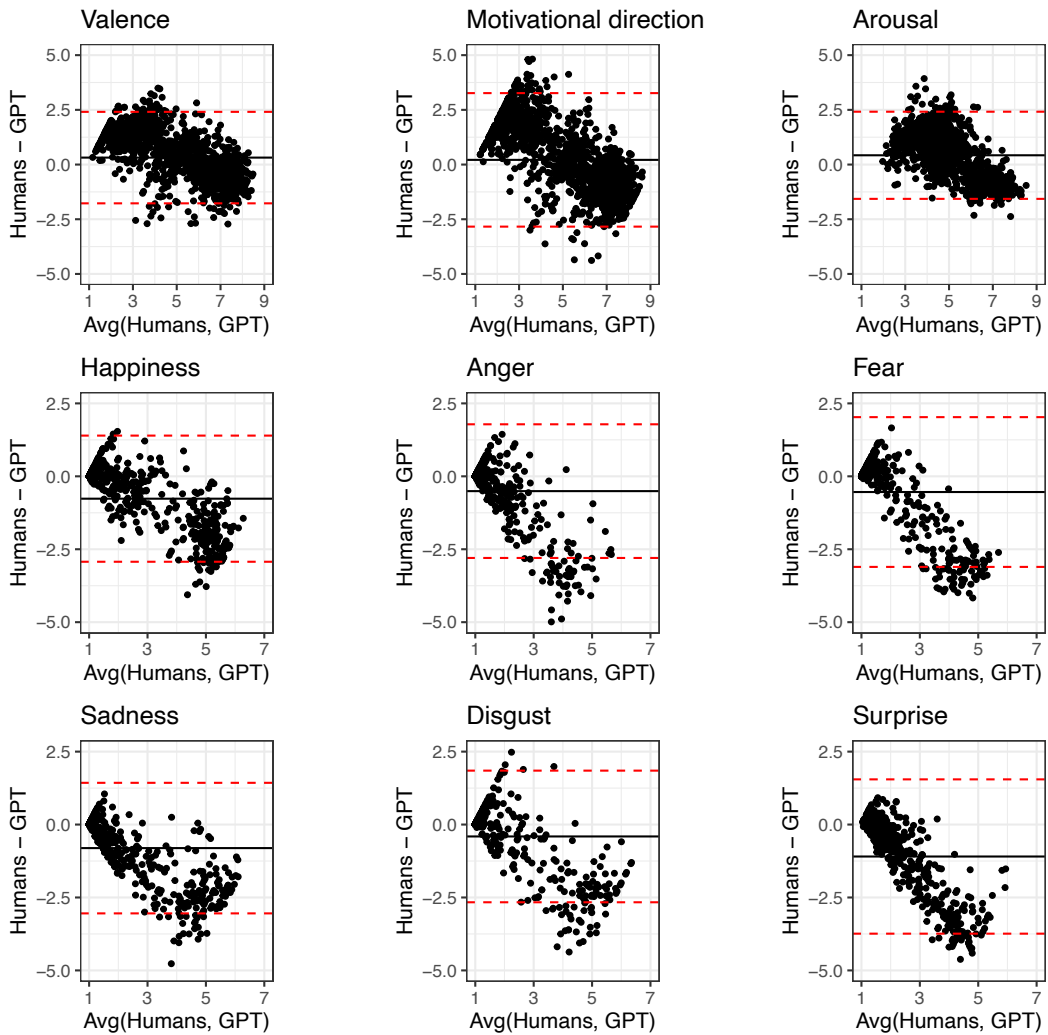

**Figure 3.** Bland-Altman plots highlighting systematic biases in GPT ratings for each affective dimension (Valence, Motivational direction and Arousal) and discrete emotion (Happiness, Anger, Fear, Sadness, Disgust and Surprise). The x-axis represents the average between Humans and GPT ratings, while the y-axis represents their difference. Solid lines represent the mean difference: values below zero indicates that GPT generally overestimates the ratings for basic discrete emotions. The markedly negative trends in the data further indicate that GPT tends to overestimate the human emotional ratings especially when stimuli are associated with strong reactions (i.e., towards the "*Very much*" bound in the 7-point Likert scale used to rate the basic emotions).